# Phase Transitions with Structured Sparsity

Huiguang Zhang[1], Baoguo Liu[2,*]

## Abstract

In the field of signal processing, phase transition phenomena have recently attracted great attention. Donoho's work established the signal recovery threshold using indicators such as restricted isotropy (RIP) and incoherence and proved that phase transition phenomena occur in compressed sampling. Nevertheless, the phase transition phenomenon of structured sparse signals remains unclear, and these studies mainly focused on simple sparse signals. Signals with a specific structure, such as the block or tree structures common in real-world applications, are called structured sparse signals. The objectives of this article are to study the phase transition phenomenon of structured sparse signals and to investigate how structured sparse signals affect the phase transition threshold. It begins with a summary of the common subspace of structured sparse signals and the theory of high-dimensional convex polytope random projections. Next, the strong threshold expression of block-structured and tree-structured sparse signals is derived after examining the weak and strong thresholds of structured sparse signals.

## 1 Introduction

Phase transitions are common in various fields such as physics, chemistry, materials science, earth sciences, ecology and economics[1][2][3]. In the area of signal processing theory, Donoho's pioneering research highlighted the presence of phase transitions in convex optimization problems related to linear programming, including applications in compressed sensing(CS) and error correction[4]. Several important theories have emerged from Donoho's findings; However, there are still unresolved issues related to phase transitions in structured sparse signals[5][6][7]. A more comprehensive examination of this topic is presented below.

The recoverability of compressed signals is a central tenet of CS, and researchers have investigated this property by measuring three key properties of the matrix: the RIP, incoherence, and phase transitions [8][9] . The RIP represents a fundamental criterion for evaluating measurement matrices and thus ensures the integrity of sparse vector geometry. Incoherence is a measure of the discrepancy between the columns of a matrix. A lower value indicates higher accuracy in signal reconstruction. Recent research has attempted to increase the efficiency of CS by optimizing these properties for specific applications[10][11]. A comprehensive study of phase transitions in CS (represented in the ($\delta$, $\rho$) plane) by Donoho and Tanner has significantly expanded our understanding of the recovery threshold[9]. This theoretical framework has been shown to be applicable to a variety of random measurement sets and extends beyond the specific matrix structure. Compared to the RIP and consistency methods, phase transition theory provides a more comprehensive insight into the recovery threshold and is in close agreement with empirical observations [12]. Several important theories have emerged from Donoho's findings, Amelunxen et al. provided a comprehensive analysis of phase transitions in convex optimization problems and showed that the locations of these transitions are determined by geometric invariants, extending the theoretical framework established by Donoho and Tanner[6]. Further research has also linked phase transitions to statistical decision theory. Zhang's work established a connection between phase transitions in convex programs and minimax risk curves in denoising problems[13], and suggests that the phase transition curve is consistent with performance metrics in various linear inverse problems. This connection highlights the relevance of phase transition theory for practical applications, particularly in signal processing and statistical analysis. Wang highlighted the implications of the Donoho-Tanner phase transition in the context of Lasso regression, showing that the performance of regression models can be sharply affected by the number of observations relative to the transition point[14]. This emphasizes the critical nature of understanding phase transitions for optimizing algorithm performance in real-world applications. Their work confirmed Donoho and Tanner's conjectures and suggested that phase transition behavior is consistent across different types of random projections. Not with standing the considerable advances made in the field of CS theory and the increasing sophistication of structured sparse signal theory[15][16]. However, with regard to structured sparse signals, a comprehensive study is still required to determine with the following question:

**whether analog phase transitions exist in the context of structured sparse signals? Furthermore, if such transitions exist, what is the threshold of their phase transition?**

The answers to the above questions will be of great help in gaining a deep understanding of the properties of structured sparse signals, improving the efficiency of the signal compression and reconstruction process, and ultimately improving the performance of existing algorithms[17].This paper aims to address this research gap and enhance our comprehension of phase transition phenomena in structured sparse signals. By examining the nexus of phase transition theory and structured sparsity, we aspire to contribute to the advancement of signal processing theory and its practical applications. The subsequent sections will delve into the current research status, the methodologies employed to study phase transition in structured sparse signals, and the prospective impact on the field of signal processing.

## 2 Theoretical Framework

Before embarking on our main undertaking, it is imperative to conduct a concise study of the random projection of high-dimensional convex polytope and the common subspace theory of structured sparse signals. This will help readers gain a deeper understanding of our work.

### 2.1 High-dimensional convex Polytope and Random Projection

The random projection of hyper polytope is an important branch of research in convex polytope theory. Due to space limitations and the research direction of this article, we will not explore the entire content of convex polytope theory in depth. Researchers interested in this topic are recommended to consult the monograph "Convex Polytopes"[18]. Co-authored by experts in the field such as Volker Kaibel, Victor Klee, Günter M. Ziegler, and Geoffrey C. Shephard, this book provides a comprehensive introduction to the research progress, classification, methods, and key achievements of convex polytopes. Among the numerous higher dimensional geometries, regular polytopes stand out for their favorable properties resulting from their inherent symmetries. In fact, a significant portion of research, including that related to phase transition theories, is based on the study of the random projection behavior of regular polytopes. As already mentioned, well-known examples include the regular simplex, the n-cube and the regular n-cross polytope. These fundamental objects are of great importance for the study of the geometric and combinatorial properties of polytopes. Research on the projection of convex hyper polytopes was advanced based on the research of scientists such as Affentranger and Schneider[19], Goodman and Pollack[20], Baryshnikov & Vitale[21], Böröczky and Henk[22], Vershik and Sporyshev[23], Donoho and Tanner [9]. Their contributions have laid the foundation for understanding how these projections influence the geometric properties and combinatorial structures of polytopes and have led to new insights in both theoretical and applied mathematics. To facilitate a more intuitive understanding of the cross polytope, we illustrate the three-dimensional and four-dimensional cross polytopes in Figure 1. It should be noted that the four-dimensional cross polytope is actually a three-dimensional projection.

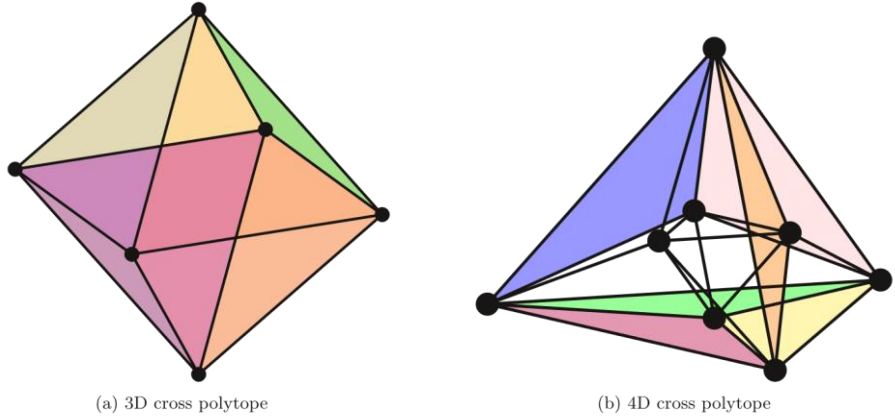

(a) 3D cross polytope  (b) 4D cross polytope

Figure. 1 3-dimensional and 4-dimensional cross-polytope

Affentranger & Schneider made significant contributions to the study of random projections of polytopes[19]. Their work laid the foundation for understanding how the geometric properties of N-dimensional polytopes change when they are projected onto lower-dimensional n subspaces. Let $Q$ be a polytope in $R^N$ and $A: R^N \to R^n$ a random ortho-projection, uniformly distributed on the Grassmannian manifold. Affentranger and Schneider proposed an effective expression for the anticipated quantity of faces of $AQ$ :

$$Ef_k(AQ) = f_k(Q) - 2\sum_{\ell}{}' \sum_{F \in F_k(Q)} \sum_{G \in F_\ell(Q)} \beta(F,G)\alpha(G,Q) \quad \text{(Equation 1)}$$

Where $f_k(Q)$ denotes the set of $k$-faces of Q, each F is a subface of G, and $\sum'$ denotes the sum over $l = n + 1, n + 3, \cdots$ where n < l < N , $\beta(F,G)$ is internal angle of $F$ and $G$, $\alpha(G,Q)$ is the external angle of $G$ and $Q$ . This is particularly important for analyzing the expected number of k-surfaces in such projections[19][22]. The core of their research is to derive a formula for the expected number of k-faces of an orthogonal projection of an n-dimensional polytope onto a randomly selected d-dimensional linear subspace. This formula is crucial for predicting how a polytope's structure will be preserved or changed during projection.

$$E(f_k(\Pi_n T^N)) \sim \frac{2^n}{\sqrt{n}} \binom{n}{k+1} \beta(T^k, T^{n-1})(\pi \ln N)^{\frac{n-1}{2}}, N \to \infty. \quad \text{(Equation 2)}$$

Where $T^N$ is N-dimension simplex, and $(\pi \ln N)^{(n-1)/2}$ as the closed form of internal angle $\alpha(G,Q)$ of $G$ and

$Q$. The projections considered by Affentranger & Schneider are onto subspaces chosen with an isotropic distribution. This means that the subspaces are selected uniformly at random, ensuring that the projections are unbiased and representative of all possible orientations [22]. This approach allows for a comprehensive analysis of how dimensionality affects the combinatorial properties of polytopes, revealing insights into their geometric behavior under various transformations. Howerver, to Böröczky and Henk, This formula still involves the "unknown" external angles $\beta(T^k, T^{d-1})$. To determine the "unknown" term of the interior angle of a polyhedron in (1), Böröczky and Henk built on the work of Daniel[24] and Baryshnikov and Vitale[21] to establish the following formula for the interior angle

$$\beta(T^k, T^{n-1}) = \frac{(k+1)^{\frac{n-k-2}{2}} e^{\frac{n-3k-3}{2}}}{\sqrt{2}^{n-k} \sqrt{\pi}^{n-k-1} d^{\frac{n-k-2}{2}}} \cdot \left(1 + O\left(\frac{k^2+1}{n}\right)\right). \quad \text{(Equation 3)}$$

This generalization of Daniels formula broadens the applicability of the original work to include more complex geometric structures[22]. based the above work, Böröczky and Henk proposed a novel complete equation of expected number of random projected regular n-cube and cross polytopes.
For fixed $k \in \mathbb{N}$, if $n/k^2 \to \infty$ and $N/n \to \infty$ then

$$E(f_k(\Pi_n T^N)) \sim E(f_k(\Pi_n C^N)) \sim \frac{\sqrt{2}^{n+k} \sqrt{\pi}^{-k} (k+1)^{\frac{n-k-2}{2}} e^{\frac{n-3k-3}{2}}}{(k+1)! \, n^{\frac{n-3k-3}{2}}} \cdot (\ln N)^{\frac{n-1}{2}}. \quad \text{(Equation 4)}$$

This result not only highlights the importance of interior angles for understanding the geometry of high-dimensional spaces, but also opens up opportunities for further research on the probabilistic behavior of such geometric configurations. In addition to the theoretical developments, Böröczky and Henk performed numerical calculations to illustrate the expected number of k-surfaces of the orthogonal projections. These calculations help validate the theoretical results and provide practical examples of the concepts discussed in the paper. Böröczky and Henk also investigated the relationship between the expected number of k-faces of the projection of a polytope and the convex hull of a standard Gaussian sample. This connection enables a probabilistic interpretation of the geometric properties of polytopes and further enriches the theoretical framework of Affentranger & Schneider [22]. Vershik and Sporyshev studied the asymptotic properties of the average number of faces in random polytopes by determining the spherical volume of a particular type of cone using the Grassmann angle[23]. Using Laplace integration and the saddle point method within integral geometry, they calculated the spherical volumes of both expanded and contracted orthotopes. They studied the asymptotic behavior of the expected number of faces of a given dimension remaining after random projections as the dimension approaches infinity while the number of faces increases proportionally.

$$f_k(\Pi_n T^N) = \begin{cases} \binom{N}{k+1}(1 + o(1)) & \text{when } \alpha > \varphi(\varepsilon), \\ \binom{N}{k+1} O(1) & \text{when } \alpha = \varphi(\varepsilon), \\ N^{-\frac{1}{2}} \exp(NS) O(1) & \text{when } \alpha < \varphi(\varepsilon), \end{cases} \quad \text{(Equation 5)}$$

Their findings implicitly reveal the phase transition behavior occurring when both the dimension and the number of faces rise in tandem, marking the first such discovery in this area. This transition suggests a critical threshold where the geometric properties of polytopes undergo significant changes, leading to new insights into their combinatorial structure and potential applications in higher-dimensional analysis. Han and Ren studied random projections of geometric sets, including convex cones, has been explored by researchers like those in Context, who provided kinematic descriptions for Gaussian random projections[25].Reitzner, Schuett, and Werner investigated the convex hull of random points on the boundary of a simple polytope, deriving asymptotic formulas for expected geometric characteristics[26].These findings not only enhance our understanding of polytopes but also open avenues for further research into their applications in optimization and data analysis, where the interplay between geometry and probability plays a crucial role.

## 2.2 Union of Subspace and Structured Block Sparsity

Although CS has made significant progress in sampling at the Nyquist rate, in practical applications, it is still possible to further reduce the required number of measurements and ensure robust signal recovery by using signal models that are more aligned with reality [27]. The key lies in adopting more realistic structurally sparse models to explore the interdependencies among signal coefficients, thus designing more efficient algorithms. For example, modern wavelet image encoders not only take advantage of the fact that a large number of wavelet coefficients are small but also consider the unique tree-like structure presented by the values and positions of large coefficients[28]. While in sparse multi-band scenarios, the main components often exhibit block-wise distributions across sparse multiband signal[17]. Model-based CS, grounded in the theory of union subspace sampling(UoS), has effectively achieved these goals[16]. This approach allows for the reconstruction of signals with fewer measurements, significantly enhancing efficiency and performance in various applications, including image processing and telecommunications. Figure 2 shows two predominant forms of structured sparse signals: block sparsity and tree sparsity. Block sparsity is simpler because non-zero elements are clustered in blocks. In contrast, tree sparsity is characterized by its total length, which is the sum of the lengths of all child nodes. Furthermore, due to the discontinuities of the signal, the tree structure typically manifests itself as a connected structure at discontinuity points.

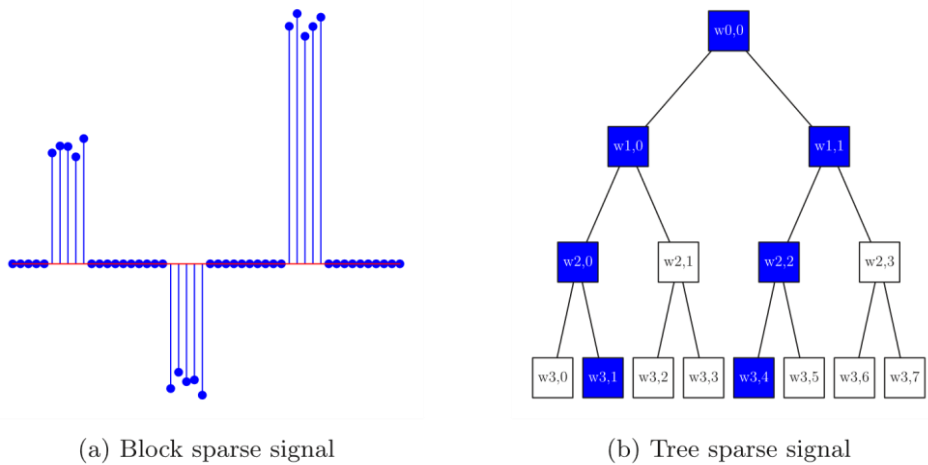

(a) Block sparse signal   (b) Tree sparse signal

Fig. 2  Structured Block Sparsity

The UoS model is a generalization of the standard sparsity assumption, allowing signals to lie in a union of subspaces rather than a single vector space[27]. This model is particularly useful in CS, where it helps in reducing the number of samples required for signal recovery[29]. The concept of block sparsity is integrated into the UoS model to enhance recovery performance. This involves arranging nonzero coefficients in blocks, which can be exploited to improve the efficiency of signal recovery algorithms [30]. The researchers derive sample bound formulas that provide conditions under which reliable subspace recovery is guaranteed. These formulas are particularly relevant in the presence of noise, where they demonstrate that fewer measurements are required compared to traditional methods using the restricted isometry property [15][29].

The principles of UoS and structured block sparsity have been applied to compressive video sampling, where they help in achieving better recovery performance and reducing computational complexity[31][32]. The integration of structured sparsity into tensor representation allows for the preservation of multidimensional signal structures, enhancing sampling efficiency and recovery performance [33].

While the work of Wimalajeewa, Eldar, and Varshney provides a robust framework for understanding and applying union of subspaces and structured block sparsity, other researchers like Yong Li and colleagues have also contributed to this field by developing models that incorporate data-driven subspaces for compressive video sampling. These models further enhance the recovery performance by adapting to the varying statistics of multidimensional signals [31][34].

## 3 Phase Transition of model-based Compressible Signals

### 3.1 Structured sparse signal phase transition theory

As already mentioned, Affentranger and Böröczky, *et.al*, have derived a closed asymptotic expression for the expected number of faces of general convex polyhedral projections using methods of integral geometry and spherical integral geometry, particularly for regular hypercubes and regular cross polytopes[19][21]. Vershik and Sporyshev studied the asymptotic behavior of the spherical volume of expanded and contracted orthogonal bodies using the Laplace integral and the saddle point method, and indirectly revealed the phase transition properties of changing the number of faces of randomly projected convex polytopes in proportionally expanding high-dimensional spaces. Based on these geometric findings, Donoho and Tanner applied the theory to the field of signal processing and proposed phase transition theory in compressed sampling, distinguishing between the concepts of strong and weak thresholds. The theory suggests that there is a critical turning point in the phase transition phenomenon of compressed sensing, the existence of which determines the success or failure of sparse signal recovery, analogous to the property transition when the dimension of convex polyhedra changes[4]. However, current research on phase transition theory focuses primarily on simple, sparse models. Although it is important to understand the compression and reconstruction processes of structured sparse signals and to develop new algorithms, there are no relevant reports to date.

Our proof is grounded in the innovative research methodologies pioneered by to Böröczky and Henk, Donoho and Tanner[9][21]. First, we will review the model the expectation of the remaining k-faces via the simple sparse signal after being randomly projected onto a high-dimensional polytope, and construct the analogical model of structured sparsity. To the (1), the discrepancy between the anticipated count of faces of the projected polytope $AQ$ and the inevitably greater count of faces of the original polytope $Q$ can be articulated as follows:

$$\Delta(k,n,N;Q) := f_k(Q) - Ef_k(AQ) = 2 \sum_{\ell}' \sum_{F \in \mathcal{F}_k(Q)} \sum_{G \in \mathcal{F}_\ell(Q)} \beta(F,G)\alpha(G,Q). \quad \text{(Equation 6)}$$

The total here encompasses the external angles formed by the original polytope Q and its subfaces G, denoted as $\beta(F,G)$ and is multiplied by the total of all internal angles between each specific subface G and its corresponding faces F, represented as $\alpha(G,Q)$.

As articulated by Donoho in his seminal works, while his approach diverges from the literature[9]. it remains fundamentally anchored in the insightful observations made by Bróczky and Henk[21] regarding specific characteristics of cross polytopes. This study not only acknowledges these foundational properties but also seeks to expand upon them. However, it is crucial to recognize the distinctions between structured sparsity and simple sparsity. Therefore, a meticulous examination of the original model is imperative, allowing us to identify and implement essential modifications that will enhance its relevance and effectiveness when applied to structured sparse signals. This careful analysis will ensure that our approach is robust and tailored to the complexities inherent in structured sparsity, thereby advancing the field further. Although the research by Donoho et al. encompasses the issues of random projections of the simplex and the cross-polytope in CS corresponding to linear programming problems, the focus of this study is primarily on the model of CS for structured sparse signals. Therefore, this paper will mainly investigate the issue of random projections of the cross-polytope, that is, we will specialize from the general case Q to $C^N$

1. To a high dimension polytype Q , or more specific in our context, a cross polytype $C^N$, the following statements about a face $F$ of $C^N$ are equivalent:
a) $AF$ is a face of $AC^N$
b) Let $y_F = \chi_F$ ,Then $\chi_F$ is the $L_1$ regular solution of the instance of $\mathcal{P}$ defined by $y_F, A$.
Despite the additional prior constraints compared to simple sparse signals, the compression and reconstruction of structured sparse signals fundamentally remain within the realm of linear programming under L1 regularization. Consequently, the aforementioned discussions remain applicable in the context of structured sparse signals.

2. With simple sparsity , There are $2^{k+1}\binom{N}{k+1}$ $k$-faces of the cross polytope in an $N$-dimensional space. while the structured sparsity framework introduces additional constraints that modify this count, so there will be only $2^{k+1}m_k$ $k$-faces of the cross polytope, where $m_k$ represents the number of valid configurations that adhere to the imposed structure.

3. For $l > k$ and specific $k$-faces ,there are $2^{\ell-k}\binom{N-k-1}{\ell-k}$ $\ell$ - faces of cross polytopes that can be derived from the existing $k$-faces, reflecting the hierarchical nature of the structured sparsity model. For a given $k$-face, since its geometric properties do not change, the properties still apply to structured sparse signals.

4. The faces of cross polytope are all simplices, and the internal angle $\beta(F,G) = \beta(T^k, T^\ell)$ , The external angle $\alpha(F^\ell, C^N)$ is the same for all $\ell$ - faces , For reasons similar to the second article, this article still applies.

From the above discussion, we can conclude the following:
for simple sparsity, the number of $k$-faces with cross polytype is:

$$f_k^{smp}(C^N) = 2^{k+1}\binom{N}{k+1} \quad \text{(Equation 7)}$$

for structured sparsity, , the number of $k$-faces with cross polytype is:

$$f_k^{str}(C^N) = 2^{k+1}m_k \quad \text{(Equation 8)}$$

Here we say a sequence of triples $((k_n, n, N_n): n = n_0, n_0 + 1, \ldots)$ is growing proportionally when there are $\delta \in (0,1)$ and $\rho \in (0,1)$ :

$$\frac{k_n}{n} \to \rho, \frac{n}{N_n} \to \delta, n \to \infty. \tag{Equation 9}$$

Subscripts n is omitted on k and N unless they are absolutely necessary the expected discrepancy of $k$-faces via simple sparsity in random projections, for $l = n + 1, n + 3, \ldots$, can be expressed as:

$$\Delta_{\ell,n}^{smp} = 2^{\ell+1} \cdot \binom{N}{k+1}\binom{N-k-1}{\ell-k}\beta(T^k, T^\ell)\alpha(F^\ell, C^N)$$
$$= C_{\ell,n}^{smp} \cdot \beta(T^k, T^\ell) \cdot \alpha(F^\ell, C^N), \tag{Equation 10}$$

Where $C_{\ell,n}^{smp} = 2^{\ell+1} \cdot \binom{N}{k+1}\binom{N-k-1}{\ell-k}$ is the combinatorial prefactor of simple sparsity.

While, the expected discrepancy between the k-face before and after projection in the case of a sparsely structured signal is

$$\Delta_{\ell,n}^{str} = 2^{\ell+1} \cdot m_k \binom{N-k-1}{\ell-k}\beta(T^k, T^\ell)\alpha(F^\ell, C^N)$$
$$= C_{\ell,n}^{str} \cdot \beta(T^k, T^\ell) \cdot \alpha(F^\ell, C^N), \tag{Equation 11}$$

Where $C_{\ell,n}^{str} = 2^{\ell+1} \cdot m_k \binom{N-k-1}{\ell-k}$ is the combinatorial prefactor of structure sparsity.

Now the necessary elements for an investigation of the change in the number of k-faces before and after a random projection in a cross polytope with simple sparsity and structural sparsity have been established. Donoho[9] et al. proposed two thresholds, namely the weak threshold and the strong threshold, which correspond to simple sparse signals. These thresholds are based on the criterion of the discrepancy between the remaining k-surfaces after projection and the original k-faces. There are two types of definitions: one is the direct definition and the other is the analytical definition. The direct definition is more intuitive and easier to understand; whereas the analytical definition is the important basis for deriving threshold expressions. Next, we first introduce the direct definition of simple sparsity.

There is a weak threshold function $\rho_W^{smp}: [0,1] \mapsto [0,1]$ in the proportional growth setting with $\rho < \rho_W^{smp}(\delta)$, $Ef_k^{smp}(AC^N)$ is

$$Ef_k^{smp}(AC^N) = f_k^{smp}(C^N)(1 - o(1)), 0 \leq k < \rho n, n \to \infty. \tag{Equation 12}$$

that is to say, Given $\rho > \rho_W^{smp}(\delta)$, for some $\epsilon > 0$ and some sequence $(k_n)$ with $k_n < \rho n$,

$$Ef_k^{smp}(AC^N) < f_k^{smp}(C^N)(1 - \epsilon), n \to \infty. \tag{Equation 13}$$

In other words, the fraction of faces lost after random projection is :

$$\frac{f_k^{smp}(C^N) - Ef_k^{smp}(AC^N)}{f_k^{smp}(C^{N-1})}. \tag{Equation 14}$$

Here, we also use a similar definition for the structured sparse $\rho_W^{str}$. Based on the definitions of $\Delta_{\ell,n}^{smp}$, $f_k^{smp}(C^N)$, $\rho_W^{smp}$, $\Delta_{\ell,n}^{str}$, $f_k^{str}(C^N)$, $\rho_W^{str}$, we can draw the following interesting conclusions.

$$\frac{f_k^{smp}(C^N) - Ef_k^{smp}(AC^N)}{f_k^{smp}(C^{N-1})} = \frac{f_k^{str}(C^N) - Ef_k^{str}(AC^N)}{f_k^{str}(C^{N-1})} = 2 \cdot \binom{N-k-1}{\ell-k} \cdot \beta(T^k, T^\ell)\alpha(T^\ell, C^N). \tag{Equation 15}$$

That is, for simple sparse and structured sparse cases, the proportion of facets lost after projection is the same. Different from the weak threshold, the strong threshold quantifies the probability of survival of all k-faces after a random projection in a high-dimensional polytope. Therefore, the strong threshold function can be defined as follows: To a function $\rho_S^{smp}: [0,1] \mapsto [0,1]$ in the proportional growth setting with $\rho < \rho_S^{\pm}(\delta)$,

In contrast to the weak threshold, the strong threshold quantifies the survival probability of all k-faces after a random projection in a high-dimensional polytope. Therefore, the strong threshold function can be defined as follows: To a function $\rho_S^{smp}: [0,1] \mapsto [0,1]$ in the proportional growth setting with $\rho < \rho_S^{smp}(\delta)$,

$$Ef_k^{smp}(AC^N) = f_k^{smp}(C^N) - o(1), 0 \leq k < \rho n, n \to \infty. \quad \text{(Equation 16)}$$

Therefore, the projected cross-polytope has, on average, the same number of k-faces as the standard cross-polytope for k slightly below $n \cdot \rho_S^{smp}(\delta)$. As opposed to this, if $\rho > \rho_S^{smp}(\delta)$, then $k_n < n\rho$ with.

$$f_k^{smp}(C^N) - Ef_k^{smp}(AC^N) \to \infty. \quad \text{(Equation 17)}$$

These strong thresholds have another interpretation. Consider the event "all low-dimensional faces survive projection". Again, we also use a similar definition for the structured sparse $\rho_S^{str}$. with $\rho < \rho_S^{str}(\delta)$,

$$Ef_k^{str}(AC^N) = f_k^{str}(C^N) - o(1) \quad \text{(Equation 18)}$$

From the previous discussion, it is clear that the direct definitions of the two types of thresholds are easy to understand. Specifically, the weak threshold is defined by the ratio of the number of k faces lost after random projection to the total number of k faces, while the strong threshold is defined by the absolute number of k faces lost. In this study, the weak threshold has consistency across both simple and structured sparsity types. In contrast, the strong threshold is subject to variation due to the different number of factor spaces. Despite the intuitiveness of the direct definition, it proves difficult to derive meaningful insights about thresholds. This article then examines the analytical definitions of both types of thresholds under the premise of proportional growth, with the aim of deriving expressions for these thresholds, with a particular focus on the strong threshold.

To the (10) and (11), The factors in $\Delta_{\ell,n}^{smp}$ and $\Delta_{\ell,n}^{str}$ have either exponential growth or decay. Here, To the $(C_{\ell,n}^{smp})$, $(C_{\ell,n}^{str})$ and $\beta(T^k, T^\ell)$, $\alpha(F^\ell, C^N)$, The combinatorial exponent $\Psi_{com}^{smp}$, $\Psi_{com}^{str}$, internal exponent $\Psi_{int}$ and external exponent $\Psi_{ext}$ are defined, so that, for any $\epsilon > 0$ and $n > n_0(\epsilon)$ and $\nu_{\ell,n} = \ell/N_n$ and $\gamma_{\ell,n} = k_n/\ell$.

$$N^{-1}\log(C_{\ell,n}^{smp}) \leq \Psi_{com}^{smp}(\nu_{\ell,n}, \gamma_{\ell,n}) + \epsilon,$$
$$N^{-1}\log(C_{\ell,n}^{str}) \leq \Psi_{com}^{str}(\nu_{\ell,n}, \gamma_{\ell,n}) + \epsilon. \quad \text{(Equation 19)}$$

and

$$N^{-1}\log(\beta(T^k, T^\ell)) \leq -\Psi_{int}(\nu_{\ell,n}, \gamma_{\ell,n}) + \epsilon,$$
$$N^{-1}\log(\alpha(F^\ell, C^N)) \leq -\Psi_{ext}(\nu_{\ell,n}) + \epsilon. \quad \text{(Equation 20)}$$

and the corresponding face exponents $\Psi_{face}^{smp}$, $\Psi_{face}^{str}$ is

$$N^{-1}\log(f_k^{smp}(C^N)) \leq \Psi_{face}^{smp}(\nu_{\ell,n}, \gamma_{\ell,n}) + \epsilon,$$
$$N^{-1}\log(f_k^{str}(C^N)) \leq \Psi_{face}^{str}(\nu_{\ell,n}, \gamma_{\ell,n}) + \epsilon. \quad \text{(Equation 21)}$$

uniformly in $l = n+1, n+3, \cdots$, where $l < N$. It follows that for $n > n_0$,
$$N^{-1}\log(\Delta_{\ell,n}^{smp}) \leq \Psi_{com}^{smp}(\nu_{\ell,n}, \gamma_{\ell,n}) - \Psi_{int}(\nu_{\ell,n}, \gamma_{\ell,n}) - \Psi_{ext}(\nu_{\ell,n}) + 3\epsilon,$$
$$N^{-1}\log(\Delta_{\ell,n}^{str}) \leq \Psi_{com}^{str}(\nu_{\ell,n}, \gamma_{\ell,n}) - \Psi_{int}(\nu_{\ell,n}, \gamma_{\ell,n}) - \Psi_{ext}(\nu_{\ell,n}) + 3\epsilon. \quad \text{(Equation 22)}$$

define the net exponents $\Psi_{net}^{smp}(\nu, \gamma)$ and $\Psi_{net}^{str}(\nu, \gamma)$ as:

$$\Psi_{net}^{smp}(\nu, \gamma) := \Psi_{com}^{smp}(\nu, \gamma) - \Psi_{int}(\nu_{\ell,n}, \gamma_{\ell,n}) - \Psi_{ext}(\nu_{\ell,n}),$$
$$\Psi_{net}^{str}(\nu, \gamma) := \Psi_{com}^{str}(\nu, \gamma) - \Psi_{int}(\nu_{\ell,n}, \gamma_{\ell,n}) - \Psi_{ext}(\nu_{\ell,n}). \quad \text{(Equation 23)}$$

A maximum value operator $M[\psi](\delta, \rho)$ on the rectangular interval $\nu \in [\delta, 1], \gamma \in [0, \rho]$ is defined as these dimensions increase proportionally. and note that $\nu_{\ell,n} = \ell/N_n$ and $\gamma_{\ell,n} = k_n/\ell$.

$$M[\psi](\delta, \rho) = \sup\{\psi(\nu, \gamma): \nu \in [\delta, 1], \gamma \in [0, \rho]\}. \quad \text{(Equation 24)}$$

pplying this operator to $\Psi_{net}^{smp}(\nu, \gamma)$ and $\Psi_{net}^{str}(\nu, \gamma)$ yields two maximal functions $M[\Psi_{net}^{smp}], M[\Psi_{net}^{str}]$. And this give analytic definitions for the strong threshold $\rho_S^{smp}(\delta)$ as the 'first' zero of $M[\Psi_{net}^{smp}]$, and $\rho_S^{str}(\delta)$ for $M[\Psi_{net}^{str}]$:

$$\rho_S^{smp}(\delta) = inf\{\rho: M[\Psi_{net}^{smp}](\delta,\rho) = 0, \rho \in [0,1]\}.$$
$$\rho_S^{str}(\delta) = inf\{\rho: M[\Psi_{net}^{str}](\delta,\rho) = 0, \rho \in [0,1]\}. \quad \text{(Equation 25)}$$

While the weak threshold $\rho_W^{smp}(\delta)$ and $\rho_W^{str}(\delta)$ can be defined the 'first' zero of $M[\Psi_{net}^{smp}/\Psi_{face}^{smp}]$ and $M[\Psi_{net}^{str}/\Psi_{face}^{str}]$. Now a comprehensive asymptotic analysis using the Donoho method is entirely possible. However, based on our conclusions above, the study of threshold for structured sparse signals can be simplified accordingly. Since the proportion of missing faces after projection is the same for simple and structured sparsity, there is no need to consider weak thresholds. We will not discuss strong thresholds in detail since the interior and exterior angle factors remain constant in two scenarios. Instead, we focus on the difference between the two combined exponents.

$$\Delta_{diff} = C_{\ell,n}^{str}/C_{\ell,n}^{smp} \quad \text{(Equation 26)}$$

And the net exponents of structured sparse signal can be expressed as:

$$\Psi_{net}^{str}(\nu,\gamma) := \Psi_{com}^{str}(\nu,\gamma) - \Psi_{int}(\nu_{\ell,n},\gamma_{\ell,n}) - \Psi_{ext}(\nu_{\ell,n})$$
$$= \Psi_{com}^{mp}(\nu,\gamma) + \frac{1}{N}\Delta_{diff}(\nu,\gamma) \quad \text{(Equation 27)}$$

It is worth noting that some may take an overly optimistic perspective and postulate that the constancy of the interior and exterior angle indices implies that it is sufficient to simply recalibrate the combined indices and the area indices according to the subspace number. However, this assumption oversimplifies the complexity involved. The task of accurately determining the subspace number expression of structured sparse signals is indeed a daunting challenge in itself. This complexity arises from the complicated nature of structured sparsity, which often requires sophisticated mathematical frameworks and algorithms to fully capture the nuances of signal representation and manipulation. Therefore, our focus on the combined indices will provide a more insightful understanding of the underlying dynamics and allow us to draw more meaningful conclusions about the behavior of the k-surfaces in this context.

Next, we focus on block-sparse and tree-sparse signals, which are more commonly used in applications, and study the corresponding phase transition phenomena. This investigation will address the unique properties and challenges associated with these types of signals, particularly the way they influence structures recovery performance and computational efficiency in practical scenarios. These signals are widely used in applications where data naturally forms groups or blocks, making them suitable for block sparse signal reconstruction algorithms.

### 3.2 Block structured sparse signal's phase transition

Block sparse signals, characterized by non-zero elements occurring in clusters, appear in various practical scenarios such as face recognition, ECG signal recovery, and PDE identification, ISAR Imaging[35][36][37]. Several scholars have contributed to the development of reconstruction algorithms and the corresponding Restricted Isometry Property (RIP) for block sparse signals, enhancing their recovery performance compared to traditional methods. Roy Y. He et al. Developed the Group Projected Subspace Pursuit (GPSP) algorithm, which leverages the Block Restricted Isometry Property (BRIP) for effective block sparse signal recovery[35]. Huang et al. Proposed high-order block RIP conditions for nonconvex block-sparse compressed sensing, ensuring exact recovery of block sparse signals. Melek and Khan Introduced the Block Adaptive Matching Pursuit (BAMP) algorithm, which adapts to the average correlation between the signal and measurement matrix blocks [36]. Zhang et al. Developed the Block Compressive Sampling Matching Pursuit (Block CoSaMP) algorithm, which improves performance for block sparse signals using Block RIP [38].

Despite the clear benefits of block sparse signal algorithms in terms of efficiency, accuracy, and robustness compared to traditional compressed sensing methods, research on the phase transition of block sparse signals remains limited[39]. The phase transition concept is crucial for understanding when exact recovery of sparse signals is possible; however, it has been primarily studied in standard sparse contexts and not adequately explored for block sparsity. This gap poses challenges since existing theories like RIP and minimum sampling lower bounds do not fully address the unique characteristics of block sparse structures. To address this research gap, we study the phase transition theory associated with Cevher's (K,C) block sparsity model[40], the Cevher's allows main coefficients to appear in up to C clusters of unknown size and position. Unlike the block sparsity model in [41][42], the (K,C) model doesn't require prior knowledge of coefficient cluster locations.

The (K, C)-sparse signal model $\mathcal{M}_{(K,C)}$ is defined as

$$\mathcal{M}_{(K,C)} = \left\{x \in R^{N+2} \,\bigg|\, \sum_{i=1}^{2C+1} \beta_i = N+2, \sum_{i=1}^{c} \beta_{2i} = K\right\}. \quad \text{(Equation 28)}$$

where $\beta_i$ $\beta_{2i}$ corresponds to the size of the zero sub-blocks and the non-zero sub-blocks, respectively.

The (K, C)-sparse signal model $\mathcal{M}_{(K,C)}$ is defined as

$$\mathcal{M}_{(K,C)} = \left\{ x \in R^{N+2} \middle| \sum_{i=1}^{2C+1} \beta_i = N + 2, \sum_{i=1}^{C} \beta_{2i} = K \right\}. \tag{Equation 29}$$

where $\beta_i$ $\beta_{2i}$ corresponds to the size of the zero sub-blocks and the non-zero sub-blocks, respectively.

Here we directly adopt Cevher's conclusion., the model's subspace count $\mathcal{B}_{(K,C)}$ is

$$\mathcal{B}_{(K,C)} = \binom{N+1-k}{C}\binom{k-1}{C-1}. \tag{Equation 30}$$

Where $N$ is the signal length, $k$ is the total number of non-zero elements, and C is the number of groups.

For the $\Psi_{net}^{str}$ of (K,C) block sparsity model, with (11) and (30) we have

$$\frac{1}{N}\log\left(\frac{\mathcal{B}_{(K,C)}}{\binom{N}{k+1}}\right) = \frac{1}{N}\log\left(\frac{(N+1-k)!}{C!\,(N+1-k-C)!} \cdot \frac{k!}{C!\,(k-C)!} \cdot \frac{k!\,(N-k)!}{N!}\right). \tag{Equation 31}$$

when $N \to \infty$ and $k \ll N$, the $\Delta_{diff}$ of block sparse signals can be simplified as :

$$\begin{aligned}
&\frac{1}{N}(N\log(N-k) + N\log(N+1-k) - N\log(N) - N\log(N+1-k-C) \\
&= \log\left(\frac{N-k}{N}\right) + \log\left(\frac{N+1-k}{N+1-k-C}\right) + \mathcal{O}(N) \\
&= \log\left(1 - \frac{k}{N}\right) + \log\left(1 + \frac{C}{N+1-k-C}\right) + \mathcal{O}(N).
\end{aligned} \tag{Equation 32}$$

Since $N \to \infty$ and $k \ll N$, $\frac{k}{N}, \frac{C}{N+1-k-C} \to 0$, And using the approximation, $\log(1+x) \approx x$, for small $x$, we get

$$\lim_{N\to\infty} \frac{1}{N}\left[\log\left(1 - \frac{k}{N}\right) + \log\left(1 + \frac{C}{N+1-k-C}\right)\right] = -\frac{k}{N} + \frac{C}{N+1-k-C} = \frac{C}{N} - \frac{k}{N}. \tag{Equation 33}$$

Here we make $C = \lfloor \zeta k \rfloor$. Where $\zeta \in [0,1]$. With the $k = \lfloor \rho \delta N \rfloor, C = \lfloor \zeta \rho \delta N \rfloor, \rho, \delta, \zeta \in [0,1]$, We get

$$\begin{aligned}
&M[\Psi_{net}^{str}](\delta, \rho, \zeta) \\
&= \delta\frac{1}{2}\left[2(\zeta - 1)\rho + \log \rho + \log\log z_\delta^{\pm} + \log(2e) + \mathcal{O}\left(\delta \vee \rho \log \rho \vee \frac{\log\log z_\delta^{\pm}}{\log z_\delta^{\pm}}\right)\right].
\end{aligned} \tag{Equation 34}$$

Where $z_\delta := (\delta\sqrt{\pi})^{-1}$ [].

Note that the introduction of the new term $2(\zeta - 1)\rho$ prevents the original threshold function $M[\Psi_{net}^{str}]$ from satisfying the condition of being equal to zero. Consequently, an adjustment is required. In this context, a function $\rho = r_S^{str}(\delta)$ is defined.

$$r_S^{str}(\delta) := r_S^{str}(\delta; \tau) := e^{(2(1-\zeta)\rho)}|\tau\log(\delta\sqrt{\pi})|^{-1} \tag{Equation 35}$$

Substituting (35) into (34) yields

$$M[\Psi_{net}^{str}](\delta, \rho, \zeta) = \delta\frac{1}{2}\left[\log\left(\frac{2e}{\tau}\right) + \mathcal{O}\left(\frac{\log\log z_\delta^{\pm}}{\log z_\delta^{\pm}}\right)\right], \delta \to 0. \tag{Equation 36}$$

The $\mathcal{O}\left(\frac{\log\log z_\delta^{\pm}}{\log z_\delta^{\pm}}\right)$ term tends to zero with $\delta$, Now $\tau > 2e$ so $\log(2e/\tau) < 0$, for some $\delta_2(\tau) > 0$, $M[\Psi_{net}^{str}](\delta, \rho, \zeta)$ stays negative on $0 < \delta < \delta_2(\tau)$. So for a block sparse signal, the expression of its threshold with respect to the $\rho, \delta, \zeta$ phase transition is

$$\rho = e^{(2(1-\zeta)\rho)}|2e\log(\delta\sqrt{\pi})|^{-1} \tag{Equation 37}$$

Obviously this is an implicit expression over $\rho, \delta, \zeta$, or, although it somewhat contradicts common sense, we can use $\rho, \zeta$ for $\delta$.

$$\begin{cases} \delta = (\sqrt{\pi})^{-1} e^{-\frac{e^{(2(1-\zeta)\rho)}}{2e\rho}}, 0 < \delta < (\sqrt{\pi})^{-1} \\ \delta = (\sqrt{\pi})^{-1} e^{\frac{e^{(2(1-\zeta)\rho)}}{2e\rho}}, (\sqrt{\pi})^{-1} < \delta < 1 \end{cases} \quad \text{(Equation 38)}$$

Considering the value range of $\delta$ and $\rho$, the function's special properties, it can actually be simplified to the following form

$$\delta = (\sqrt{\pi})^{-1} e^{-\frac{e^{(2(1-\zeta)\rho)}}{2e\rho}} \quad \text{(Equation 39)}$$

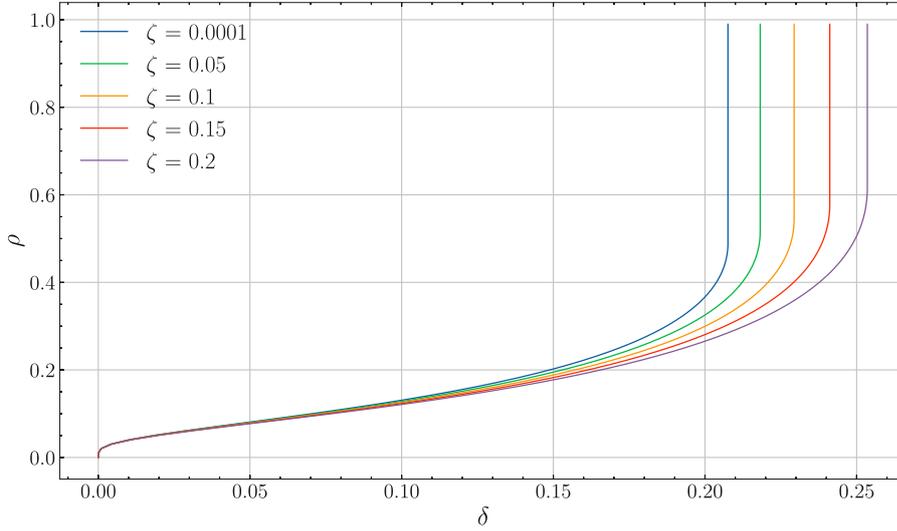

Figure. 3 Phase transition threshold for block sparse signals

Fig. 3 exhibits the trend of the strong threshold for phase transition in block sparse signals as the parameter $\zeta$ varies. It is observed that with the escalation of $\zeta$, indicative of an escalation in the number of blocks, the threshold manifests a progressively declining tendency, a observation that concurs with the findings reported in the extant literature[40].

### 3.3 Tree structured sparse signal's phase transition

Tree Sparse signals are a specific type of sparse signals where the non-zero elements are organized in a tree-like structure. These signals commonly appear in various applications such as image modeling, genetic data analysis, and CS, where the hierarchical nature of the data can be naturally represented as a tree structure[43].

Several scholars have contributed significantly to the study of reconstruction algorithms for sparse tree signals and their associated Restricted Isometry Property (RIP). Lee and Shim developed innovative algorithms, including Tree Search Matching Pursuit (TSMP) and Greedy Tree Matching Pursuit (GTMP), which employ tree search and pruning strategies to enhance sparse signal recovery[44][45] (Lee et al., 2016; Lee & Shim, 2014). Wang and Wu proposed the Exact Tree Projection (ETP) algorithm in conjunction with CoSaMP to improve the reconstruction accuracy of tree-structured CS[46][47].

Despite these advantages in efficiency, accuracy, and robustness compared to traditional compressed sensing algorithms, research on the phase transition of tree sparse signals remains limited. The phase transition concept is essential for understanding when exact recovery is feasible; however, it has not been thoroughly investigated in the context of tree sparsity. This gap presents challenges since existing theories such as RIP and minimum sampling lower bounds do not fully account for the unique characteristics of tree sparse structures. Addressing this gap could lead to enhanced theoretical frameworks and practical applications that further refine recovery processes for tree sparse signals across various fields.

To address this research gap, we study the phase transition theory associated with Baraniuk's tree sparsity model[16], In this study, we employ Baraniuk's signaling model on wavelets. Consider a signal x of length $N = 2^I$, where I is an integer. x is expressed as follows

$$x = v_0 v + \sum_{i=0}^{I-1} \sum_{j=0}^{2^i-1} w_{i,j} \psi_{i,j} \quad \text{(Equation 40)}$$

where $v$ is the scaling function and $\psi_{i,j}$ is the wavelet function at the scale i and offset j. A wavelet transform consists of a scaling factor $v_0$ and a wavelet coefficient $w_{i,j}$, which at the scale $i, 0 < i < I-1$ and position $j, 0 \leq j \leq 2^i - 1$, x is represented by the transform matrix $x = \Psi\alpha$.

Detectors of local discontinuities are wavelet functions. Making use of wavelets' nested support property at different scales makes it clear that signal discontinuities will appear as a series of noticeable and faint wavelet coefficients along the branches that run from the wavelet tree's leaves to its roots. On the other hand, a unique set of wavelet coefficients will be produced by regions with smooth signal characteristics. There are many wavelet-based processing and compression algorithms that make extensive use of this "connected tree" property.

Define the set of -tree sparse signals a

$$T_K = \begin{cases} x = v_0 v + \sum_{i=0}^{l-1} \sum_{j=1}^{2^i} w_{i,j} \psi_{i,j}: w|_\Omega c = 0, \\ |\Omega| = K, \Omega\ forms\ a\ connected\ subtree. \end{cases} \quad \text{(Equation 41)}$$

Here we directly use Baraniuk's research results on the number $T_K$ of subspaces in sparse tree structures. The number k different connected, The number of subspaces in $T_K$ obeys $T_K \leq \frac{(2e)^k}{k+1}$ for $k < \log_2 N$ and $T_K \leq \frac{4^{k+4}}{ke^2}$ for $k \geq \log_2 N$

So, For the $\Psi_{net}^{str}$ of Tree sparsity model with $k < \log_2 N$ we have

It is therefore necessary to discuss the two cases separately. Regarding the sparsity of $k < \log_2 N$, we have

$$\Psi_{net}^{str} = \frac{1}{N} \log \left( \frac{\frac{(2e)^k}{k+1}}{\binom{N}{k+1}} \right)$$

$$= \frac{1}{N}(k\log(2e) + N\log(N-k-1) - N\log N) \quad \text{(Equation 42)}$$

$$= \frac{k}{N}\log(2e) + \log(N-k-1) - \log N$$

$$= \frac{k}{N}\log(2).$$

$$M[\Psi_{net}^{str}](\delta, \rho, \zeta)$$
$$= \delta \frac{1}{2}\left[ 2(\log(2))\rho + \log \rho + \log\log z_\delta^\pm + \log(2e) + \mathcal{O}\left(\delta \vee \rho \log \rho \vee \frac{\log\log z_\delta^\pm}{\log z_\delta^\pm}\right) \right]. \quad \text{(Equation 43)}$$

Once more, the advent of the novel term $2(\log(e))\rho$ precludes the original threshold function $M[\Psi_{net}^{str}]$ from fulfilling the requisite condition of equaling zero. It is necessary to make the requisite adjustments. In this instance, we define a function $\rho = r_S^{str}(\delta)$:

$$r_S^{str}(\delta) := r_S^{str}(\delta; \tau) := e^{\left(2\left(\log\left(\frac{1}{2}\right)\right)\rho\right)} |\tau \log(\delta\sqrt{\pi})|^{-1} \quad \text{(Equation 44)}$$

The $\mathcal{O}\left(\frac{\log\log z_\delta^\pm}{\log z_\delta^\pm}\right)$ term tends to zero with $\delta$, Now $\tau > 2e$ so $\log(2e/\tau) < 0$, for some $\delta_2(\tau) > 0$, $M[\Psi_{net}^{str}](\delta, \rho, \zeta)$ stays negative on $0 < \delta < \delta_2(\tau)$. Therefore, we get the expression for the threshold of the phase transition of the tree-like sparse signal with sparsity $K < \log_2 N, \rho, \delta, \zeta$ as

$$\rho = e^{\left(2\left(\log\left(\frac{1}{2}\right)\right)\rho\right)} |2e\log(\delta\sqrt{\pi})|^{-1} \quad \text{(Equation 45)}$$

Obviously, this is still an expression implicitly defined in terms of $\rho, \delta$. We can still use the notation $\rho$ to represent $\delta$ in reverse.

Obviously, this is still an expression implicitly defined by $\rho, \delta$. We can still use $\rho$ to represent $\delta$ in reverse.

$$\begin{cases} \delta = (\sqrt{\pi})^{-1} e^{-\frac{e^{\left(2\left(\log\left(\frac{1}{2}\right)\right)\rho\right)}}{2e\rho}}, 0 < \delta < (\sqrt{\pi})^{-1}, \\ \delta = (\sqrt{\pi})^{-1} e^{\frac{e^{\left(2\left(\log\left(\frac{1}{2}\right)\right)\rho\right)}}{2e\rho}}, (\sqrt{\pi})^{-1} < \delta < 1. \end{cases} \quad \text{(Equation 46)}$$

Similarly, the above formula can also be simplified as follows:

$$\delta = (\sqrt{\pi})^{-1} e^{\frac{e^{\left(2\left(\log\left(\frac{1}{2}\right)\right)\rho\right)}}{2e\rho}}$$

And for the case of $K \geq \log_2 N$

$$\Psi_{net}^{str} = \frac{1}{N}\log\left(\frac{\frac{4^{k+4}}{ke^2}}{\binom{N}{k+1}}\right)$$

$$= \frac{1}{N}(klog(4) + Nlog(N-k-1) - NlogN) \quad \text{(Equation 47)}$$

$$= \frac{k}{N}\log(4) + \log(N-k-1) - logN$$

$$= \frac{k}{N}(log(4) - 1).$$

$$M[\Psi_{net}^{str}](\delta,\rho,\zeta)$$
$$= \delta\frac{1}{2}\left[2(\log(4)-1)\rho + \log\rho + \log\log z_\delta^\pm + \log(2e) + \mathcal{O}\left(\delta \vee \rho\log\rho \vee \frac{\log\log z_\delta^\pm}{\log z_\delta^\pm}\right)\right]. \quad \text{(Equation 48)}$$

we define a function $\rho = r_S^{str}(\delta)$ as:

$$r_S^{str}(\delta) := r_S^{str}(\delta;\tau) := e^{(2\tau(1-log(4))\rho)}|log(\delta\sqrt{\pi})|^{-1} \quad \text{(Equation 49)}$$

The $\mathcal{O}\left(\frac{\log\log z_\delta^\pm}{\log z_\delta^\pm}\right)$ term tends to zero with $\delta$, Now $\tau > 2e$ so $\log(2e/\tau) < 0$, for some $\delta_2(\tau) > 0$, $M[\Psi_{net}^{str}](\delta,\rho,\zeta)$ stays negative on $0 < \delta < \delta_2(\tau)$. Therefore, we get the expression for the phase transition threshold of a tree-like sparse signal with sparsity $K \geq \log_2 N$ as

$$\rho = e^{(2(1-log(4))\rho)}|\tau log(\delta\sqrt{\pi})|^{-1} \quad \text{(Equation 50)}$$

The relationship between x and y can be expressed as

$$\begin{cases} \delta = (\sqrt{\pi})^{-1} e^{-\frac{e^{(2(1-log(4))\rho)}}{2e\rho}}, 0 < \delta < (\sqrt{\pi})^{-1} \\ \delta = (\sqrt{\pi})^{-1} e^{\frac{e^{(2(1-log(4))\rho)}}{2e\rho}}, (\sqrt{\pi})^{-1} < \delta < 1 \end{cases} \quad \text{(Equation 51)}$$

Similarly, the above formula can also be simplified as follows:

$$\delta = (\sqrt{\pi})^{-1} e^{\frac{e^{(2(1-log(4))\rho)}}{2e\rho}} \quad (52)$$

Behaviors that exceed the threshold can also be studied and proven in a similar way as described above.

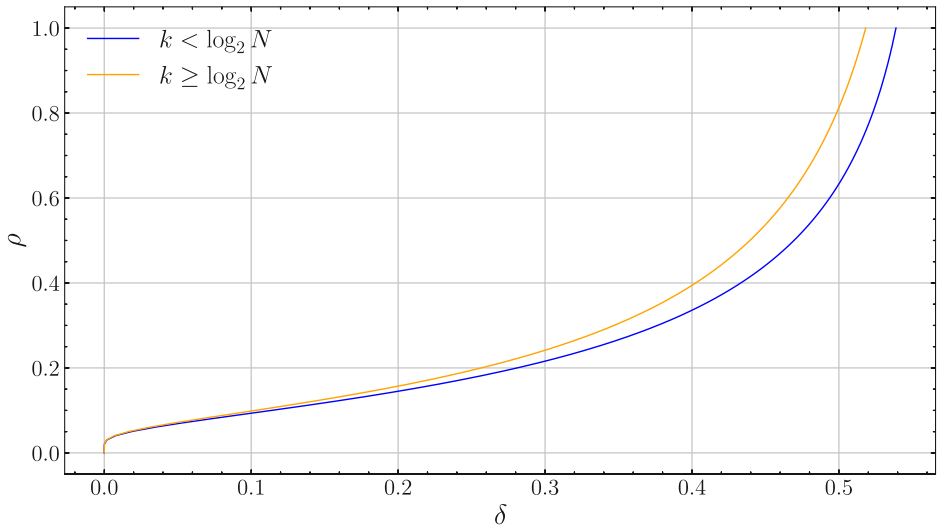

Figure. 4 Phase transition threshold for tree sparse signals

Fig. 4 presents the trend of the strong phase transition threshold corresponding to the counts of the two sub-models under various sparsity conditions. Since the counting method $K < \log_2 N$ is rather conservative, its corresponding threshold estimates are also relatively conservative, a result that is consistent with findings from existing literature[16].

## 4 Discussion

Under both simple sparsity and structured sparsity, the percentage of faces lost following a random projection is known as the weak threshold. In other words, the percentage of faces lost following a random projection below the weak threshold is constant, independent of the signal structure. After the random projection, the probability that all k faces will survive is known as the strong threshold. Because of the impact of structured sparsity, it might alter. This is because the definition of the strong threshold is altered by structured sparsity, which also influences the number of subspaces. Because structured sparse signals have fewer subspaces, they have a lower strong threshold. At the same measurement space dimension, this suggests that structured sparse signals are more likely to recover accurately. The number of blocks, sparsity, and signal length all affect the implicit expression for the phase transition threshold of a block-structured sparse signal. An implicit expression that is dependent on the sparsity and signal length is also the phase transition threshold of a tree-structured sparse signal. In contrast to simple sparse signals, block-structured sparse signals appear to exhibit more complex phase transition behavior. There are many variables that affect it, and it is hard to put into a straightforward formula.

## 5 Conclusion

To determine the strong threshold expressions for block-structured and tree-structured sparse signals, we investigate the phase transition phenomenon of structured sparse signals. It was discovered that while the strong threshold's absolute value dropped as a result of the subspace number change, the weak threshold stayed constant. However, the lack of knowledge regarding the phase transition's width limits this investigation. By closely analyzing the phase transition's width and the variations among various structured sparse signal types, more research could close this knowledge gap.

## References

[1] Salinas, S. R. A. (2001). Phase transitions and critical phenomena: classical theories. In Introduction to Statistical Physics (pp. 235-256).

[2] Papon, P., Leblond, J., & Chassande, M. (2002). Thermodynamics and statistical mechanics of phase transitions. In The Physics of Phase Transitions: Concepts and Applications (pp. 1-35). Springer.

[3] Hindmarsh, M., Kunz, J., & Olinto, A. (2021). Phase transitions in the early universe. SciPost Physics Lecture Notes, 2021(24).

[4] Donoho, D. L., & Tanner, J. (2009). Counting faces of randomly projected polytopes when the projection radically lowers dimension. Journal of the American Mathematical Society, 22(1), 1-53. https://doi.org/10.1090/S0894-0347-08-00600-0

[5] Hu, S.-W., Li, F., & Goldgof, D. B. (2015). Phase transition of joint-sparse recovery from multiple measurements via convex optimization. In 2015 IEEE International Conference on Acoustics, Speech and Signal Processing (ICASSP) (pp. 4409-4413). IEEE. https://doi.org/10.1109/ICASSP.2015.7178698

[6] Amelunxen, D., Lotz, M., McCoy, M. B., & Tropp, J. A. (2014). Living on the edge: Phase transitions in convex programs with random data. Information and Inference: A Journal of the IMA, 3(3), 224-294. https://doi.org/10.1093/imaiai/iau005

[7] Sun, Z., Cui, W., & Liu, Y. (2022). Phase transitions in recovery of structured signals from corrupted measurements. IEEE Transactions on Information Theory, 68(7), 4837-4863. https://doi.org/10.1109/TIT.2022.3153204

[8] Candes, E. J. (2008). The restricted isometry property and its implications for compressed sensing. Comptes rendus. Mathematique, 346(9-10), 589-592. https://doi.org/10.1016/j.crma.2008.03.014

[9] Donoho, D. L., & Tanner, J. (2009). Counting faces of randomly projected polytopes when the projection radically lowers dimension. Journal of the American Mathematical Society, 22(1), 1-53. https://doi.org/10.1090/S0894-0347-08-00600-0

[10] Chen, Y., & Lin, S. (2021). Preconditioning with RIP improvement for compressed sensing systems in total noise environment. Signal Processing, 179, 107765. https://doi.org/10.1016/j.sigpro.2020.107765

[11] Bertocco, M., Frigo, G., & Narduzzi, C. (2015). High-accuracy frequency estimation in compressive sensing-plus-DFT spectral analysis. In 2015 IEEE International Instrumentation and Measurement Technology Conference (I2MTC) Proceedings (pp. 1-6). IEEE. https://doi.org/10.1109/I2MTC.2015.7149768

[12] Bertsimas, D., & Van Parys, B. (2020). SPARSE HIGH-DIMENSIONAL REGRESSION. The Annals of Statistics, 48(1), 300-323. https://doi.org/10.1214/18-AOS1772

[13] Zhang, Huan, Yulong Liu, and Hong Lei. "Phase Transition of Convex Programs for Linear Inverse Problems with Multiple Prior Constraints." *arXiv preprint arXiv:1801.00965* (2018).

[14] Wang, Hua, et al. "The complete Lasso tradeoff diagram." *Advances in Neural Information Processing Systems* 33 (2020): 20051-20060.

[15] Blumensath, T., & Davies, M. E. (2009). Sampling theorems for signals from the union of finite-dimensional linear subspaces. IEEE Transactions on Information Theory, 55(4), 1872-1882. https://doi.org/10.1109/TIT.2009.2012898

[16] Baraniuk, R. G., Cevher, V., Duarte, M. F., & Hegde, C. (2010). Model-based compressive sensing. IEEE Transactions on Information Theory, 56(4), 1982-2001. https://doi.org/10.1109/TIT.2010.2040894

[17] Eldar, Y. C., Kuppinger, P., & Bolcskei, H. (2010). Block-sparse signals: Uncertainty relations and efficient recovery. IEEE Transactions on Signal Processing, 58(6), 3042-3054. https://doi.org/10.1109/TSP.2010.2043134

[18] Grünbaum, Branko, et al. *Convex polytopes.* Vol. 16. New York: Interscience, 1967.

[19] Affentranger, F., & Schneider, R. (1992). Random projections of regular simplices. Discrete & Computational Geometry, 7(3), 219-226. https://doi.org/10.1007/BF02187840

[20] Goodman, J. E., & Pollack, R. (1986). There are asymptotically far fewer polytopes than we thought. In Discrete & Computational Geometry (pp. 127-129). Springer.

[21] Baryshnikov, Y. M., & Vitale, R. A. (1994). Regular simplices and Gaussian samples. Discrete & Computational Geometry, 11(2), 141-147. https://doi.org/10.1007/BF02574011

[22] Böröczky, K. J., & Henk, M. (1999). Random projections of regular polytopes. Archiv der Mathematik, 73, 465-473. https://doi.org/10.1007/s000130050390

[23] Vershik, A. M., & Sporyshev, P. V. (1992). Asymptotic behavior of the number of faces of random polyhedra and the neighborliness problem. Selecta Math. Soviet, 11(2), 181-201.

[24] Rogers, C. A. "Packing and Covering, Cambridge University Press." *London, England* (1964).

[25] Han, Q., & Ren, H. (2022). Gaussian random projections of convex cones: approximate kinematic formulae and applications. arXiv preprint arXiv:2212.05545.

[26] Reitzner, M., Schütt, C., & Werner, E. M. (2023). The convex hull of random points on the boundary of a simple polytope. Discrete & Computational Geometry, 69(2), 453-504. https://doi.org/10.1007/s00454-022-00414-6

[27] Lu, Y. M., & Do, M. N. (2008). A theory for sampling signals from a union of subspaces. IEEE Transactions on Signal Processing, 56(6), 2334-2345. https://doi.org/10.1109/TSP.2008.916064

[28] La, C., & Do, M. N. (2005). Signal reconstruction using sparse tree representations. In Wavelets XI (Vol. 5914, pp. 1-12). SPIE. https://doi.org/10.1117/12.632901

[29] Wimalajeewa, T., Eldar, Y. C., & Varshney, P. K. (2015). Subspace recovery from structured union of subspaces. IEEE Transactions on Information Theory, 61(4), 2101-2114. https://doi.org/10.1109/TIT.2015.2393570

[30] Eldar, Y. C., & Mishali, M. (2009). Block sparsity and sampling over a union of subspaces. In 2009 16th International Conference on Digital Signal Processing (pp. 318-327). IEEE. https://doi.org/10.1109/ICDSP.2009.5204474

[31] Li, Y., Dai, W., & Xiong, H. (2019). Scalable structured compressive video sampling with hierarchical subspace learning. IEEE Transactions on Circuits and Systems for Video Technology, 30(10), 3528-3543. https://doi.org/10.1109/TCSVT.2019.2911475

[32] Li, Y., Dai, W., & Xiong, H. (2015). Subspace learning with structured sparsity for compressive video sampling. In 2015 Data Compression Conference (DCC) (pp. 159-168). IEEE. https://doi.org/10.1109/DCC.2015.7104043

[33] Li, Y., Dai, W., & Xiong, H. (2016). Compressive tensor sampling with structured sparsity. In 2016 Data Compression Conference (DCC) (pp. 201-210). IEEE. https://doi.org/10.1109/DCC.2016.7499539

[34] Li, Y., Dai, W., Xiong, H., & Hsu, W. (2017). Structured sparse representation with union of data-driven linear and multilinear subspaces model for compressive video sampling. IEEE Transactions on Signal Processing, 65(19), 5062-5077. https://doi.org/10.1109/TSP.2017.2724976

[35] He, R. Y., Liu, H., & Liu, H. (2024). Group Projected Subspace Pursuit for Block Sparse Signal Reconstruction: Convergence Analysis and Applications. arXiv preprint arXiv:2407.07707.

[36] Melek, M. (2023). Efficient Block-sparse Signal Recovery with Application to ECG Compression. In 2023 40th National Radio Science Conference (NRSC) (pp. 1-4). IEEE. https://doi.org/10.1109/NRSC56532.2023.9880736

[37] Zhao, J., Lin, S., & Huang, T. (2021). Block Sparse Bayesian Recovery with Correlated LSM Prior. Wireless Communications and Mobile Computing, 2021, 9942694. https://doi.org/10.1155/2021/9942694

[38] Zhang, Xiaobo, et al. "On recovery of block sparse signals via block compressive sampling matching pursuit." *IEEE Access* 7 (2019): 175554-175563.

[39] Tanaka, Toshiyuki. "Phase Transition in Mixed ℓ2/ℓ1-norm Minimization for Block-Sparse Compressed Sensing." *2019 IEEE International Symposium on Information Theory (ISIT)*. IEEE, 2019.

[40] Cevher, Volkan, et al. "Recovery of clustered sparse signals from compressive measurements." International Conference on Sampling Theory and Applications (SAMPTA). 2009.

[41] Eldar, Yonina C., and Moshe Mishali. "Robust recovery of signals from a structured union of subspaces." IEEE Transactions on Information Theory 55.11 (2009): 5302-5316.

[42] Stojnic, Mihailo, Farzad Parvaresh, and Babak Hassibi. "On the reconstruction of block-sparse signals with an optimal number of measurements." IEEE Transactions on Signal Processing 57.8 (2009): 3075-3085.

[43] Chen, C., & Huang, J. (2012). Compressive sensing MRI with wavelet tree sparsity. In Advances in neural information processing systems (pp. 164-172).

[44] Lee, J., Choi, J. W., & Shim, B. (2016). Sparse signal recovery via tree search matching pursuit. Journal of Communications and Networks, 18(5), 699-712. https://doi.org/10.1109/JCN.2016.000124

[45] Lee, J., & Shim, B. (2014). Sparse Signal Recovery Using A Tree Search. The Journal of Korean Institute of Communications and Information Sciences, 39(12), 756-763.

[46] Wang, M., Li, Y., & Xiong, H. (2016). Reconstruction algorithm using exact tree projection for tree-structured compressive sensing. IET Signal Processing, 10(5), 566-573. https://doi.org/10.1049/iet-spr.2015.0178

[47] Cen, Y., Li, Y., & Xiong, H. (2013). Tree-Based Backtracking Orthogonal Matching Pursuit for Sparse Signal Reconstruction. Journal of Applied Mathematics, 2013, 864132. https://doi.org/10.1155/2013/864132